# Study on collimation and shielding of the back-streaming neutrons at the CSNS target

JING Han-Tao, TANG Jing-Yu, YANG Zheng

Institute of High Energy Physics, Chinese Academy of Sciences

**Abstract:** The back-streaming neutrons from the spallation target at CSNS are very intense, and can pose serious damage problems for the devices in the accelerator-target interface region. To tackle the problems, a possible scheme for this region was studied, namely a specially designed optics for the proton beam line produces two beam waists, and two collimators are placed at the two waist positions to maximize the collimation effect of the back-streaming neutrons. Detailed Monte Carlo simulations with the beams in the two different CSNS phases show the effectiveness of the collimation system, and the radiation dose rate decreases largely in the interface section. This can ensure the use of epoxy coils for the last magnets and other devices in the beam transport line with reasonable lifetimes, e.g. thirty years. The design philosophy for such an accelerator-target interface region can also be applicable to other high-power proton beam applications.

## 1. Introduction

The China Spallation Neutron Source (CSNS) is a large scientific facility under construction, mainly for multidisciplinary research on material characterization using neutron scattering techniques [1, 2]. The CSNS complex is composed of a high-power proton accelerator, a spallation target, neutron instruments, and conventional facilities, as shown in Figure 1. The 100-kW proton beam at 1.6 GeV and 65 μA with a repetition rate of 25 Hz is the design goal for the CSNS Phase One or CSNS-I. The upgrading potential to 500 kW has been reserved. For such a high-power proton accelerator, the beam loss control along the accelerator is essential for the hands-on maintenance and the good lifetime of the accelerator components. The loss level will be controlled below a level of 1 W/m for most of the proton beam line. The locations with higher beam loss or intentionally installed collimators will be treated with remote handling or local strengthened shielding. However, in the accelerator-target interface region or the last part of RTBT (Ring-to-Target Beam Transport) beam line as marked by a dashed circle in Figure 1, the main contribution to the radiation dose rate is not from proton beam loss but from the back-streaming neutrons from the target. This article presents the study about the collimation and shielding of the back-streaming neutrons.

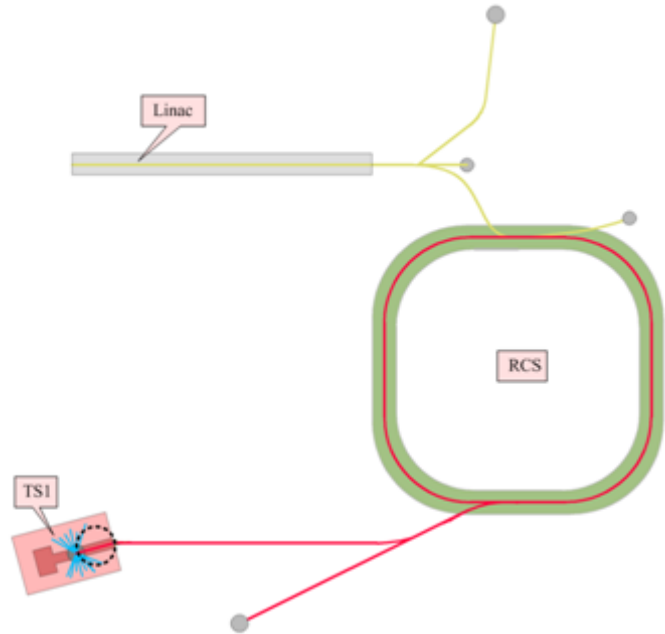

Figure 1: Schematic of CSNS layout

At CSNS, difference design schemes for the accelerator-target interface region were studied; one of them is presented here which employs step-like field magnets for uniformizing the beam spot at target and two sets of collimators at the two proton beam waists. Two measures are proposed to reduce the radiation level along RTBT by the very intense back-streaming neutrons. One is that a bending magnet of about 15° is used just before the proton beam enters the experimental building, so that only a short part of RTBT is affected by the back-streaming neutrons and the upstream part of the proton beam line is cleared from the back-streaming neutrons. In addition, the neutrons can be employed as a white neutron source for nuclear data measurements [3]. Another measure is that a collimation system just in front of the target is used to collimate the majority of the neutron flux. Although the design scheme is partially different from the one finally adopted at CSNS, it has significant advantages and can be applied in other high-power proton beam applications.

Different solutions have been studied for the treatment of back-streaming neutrons in different high-power hadron beam applications. For example, a shielding wall before the last quadrupole doublet is inserted to avoid activation of the beam line induced by the back-streaming neutrons from the target at IFMIF [4]. A collimator to collimate both the proton beam and the back-streaming neutron flux is used at SINQ [5]. At ESS [6], the proton beam will be lifted by 4.1 m by a vertical bending section from the accelerator tunnel level to the ground level of the experimental hall, where two collimators are adopted to collimate both the proton beam and the back-streaming neutrons. At SNS [7] and J-PARC [8], no such collimators are used in front of the target. At SNS, so-called T-shape tunnel with shielding material filling the unused space was designed to reduce the radiation dose rate in the RTBT; at J-PARC, the beam transport line 3NBT is protected by the collimators around the muon target that is close the spallation target and by using the T-shape-like tunnel design. When the radiation dose rate in the last section of beam transport line is too high, radiation-resistant magnets should be used, such as at SNS and J-PARC. Obviously, such magnets, usually using mineral-insulated conductor (or MIC) coils, are expensive and complicated for manufacturing when compared to the conventional epoxy coil magnets.

Therefore, the study presented here will show if conventional coil magnets will have good lifetimes in the radiation level at the CSNS, when a good collimation system is applied.

## 2. Accelerator-target interface section at CSNS

The RTBT has a total length of about 140 m and consists of three straight sections separated by two bending magnets. The first bending magnet guides the proton beam to either the second section of RTBT or the beam dump. The second bending magnet will deflect the proton beam by an angle of 15° in order to avoid the back-streaming radiation further along the RTBT tunnel. The length of the last straight section of RTBT, from the second bending magnet to the target center, is about 21 m, which is labeled by a dotted circle in Figure 1 and also called the accelerator-target interface section. It receives the highest neutron irradiation in RTBT. In this paper, we will study the radiation dose level in the section and the collimation method.

The schematic layout of the accelerator-target interface section is shown in Figure 2. There are many components in the section, including three quadrupole magnets, one dipole magnet, two sets of nonlinear magnets (each with two units), the beam diagnostics devices, the proton beam window (PBW), the vacuum devices, the collimators, and so on. One of the key issues in this interface section is the collimation of the incoming proton beam and the back-streaming neutrons. The former is to protect the target, the PBW and the moderators when the proton beam deviates from the usual setting; the latter is to reduce the radiation dose rate in the RTBT. At CSNS, the collimation system consists of two collimators in the last long drift after the last magnet. In order to enhance the collimation effect, the proton beam optics is designed to have two waists in the long drift, one in the horizontal plane and the other in the vertical plane, so that narrow apertures can be used for the collimators at the waists. The apertures of the collimators are chosen so that the power deposit of the incident proton beam halo in the collimators is below 500 W in the normal operation for the sake of natural cooling of the collimators, and they fit to the proton beams both at CSNS-I and CSNS-II.

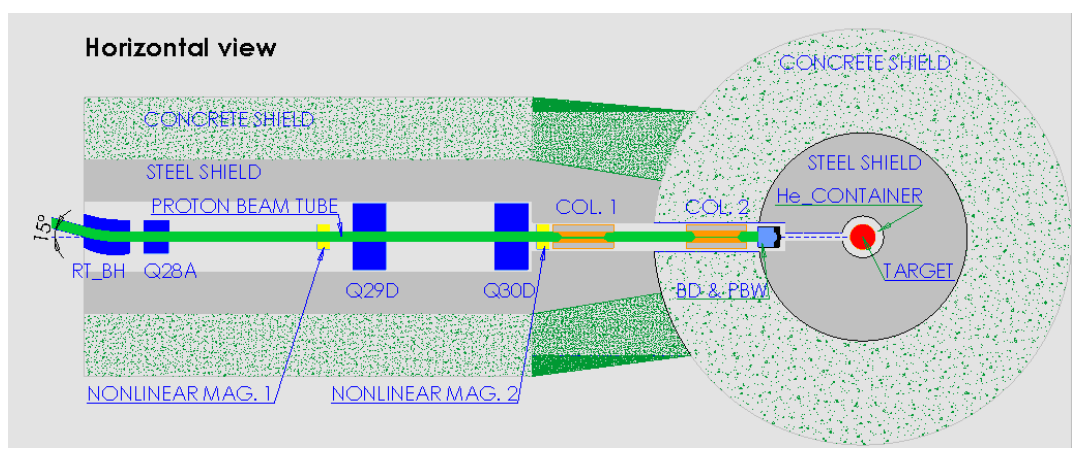

Figure 2: Schematic of the accelerator-target interface section at CSNS

## 3. Collimation method and simulation results

The optics design of the proton beam line and the multiple particle simulations are carried out by using the TRANSPORT [9] and TURTLE codes [10]. The spot uniformization at target is performed by two sets of nonlinear magnets or step-like field magnets in the last RTBT section [11]. The Monte-Carlo code FLUKA [12, 13] is used for the simulations of the neutron production in the target and the collimation effect of the back-streaming neutrons by the collimators.

The nominal beam footprint at the target defines the beam spot dimensions containing the majority of the beam upon to specific definition, e.g. about 95% for a Gaussian distribution and higher portion for more uniform beams, which is 40 mm (V) × 120 mm (H) at CSNS-I. For the upgrading phase CSNS-II, the nominal beam footprint is 60 mm (V) × 160 mm (H), and the larger footprint is required to reduce the heat density in the spallation target. There possibly exists a medium upgrading stage with the beam power of 200 kW, but the footprint will be the same as that in CSNS-I.

At the proton energy of 1.6 GeV, the collimators in copper have a length of about 1 m. Copper is considered a good material to be the proton collimator and the neutron absorber due to its good thermal conductivity, high mass density and neutron property. To benefit from the beam waists design, both collimators have four blocks (upper, lower, left and right) with each having its inner surface following the proton beam envelopes. All the back-streaming neutrons hitting the collimator blocks will be absorbed or scattered, and the scattered neutrons will be most probably absorbed or shielded by the surrounding shielding materials.

As we intend to build the collimators with fixed apertures, the design of the collimators is the balanced result between CSNS-I and CSNS-II conditions. Different operation scenarios have been considered in the simulations: (1) normal operation mode at CSNS-I and CSNS-II; (2) low-power operation mode without nonlinear magnets in use.

### 3.1 Positions of collimators

The last nonlinear magnet and proton beam window are at 7.8 m and 2.0 m from the target, respectively. Therefore, the two collimators must be in between them. As mentioned above, it is preferred that the collimators are placed at the positions of the horizontal and vertical beam waists in order to intercept the neutrons and other particles back-streaming from the target efficiently. However, the horizontal beam waist is at the last nonlinear magnets for the reason of good uniformization effect, the first collimator can be placed just after the magnets. The proton beam envelopes from the last nonlinear magnet to the target at CSNS-I and CSNS-II are shown in Figure 3. Taking into account the dimensions of the collimators, the surrounding shielding and the required installation space, the entrance locations of the two collimators are at -7.2 m and -4.5 m, respectively. Obviously, the second collimator has to be inside the target shielding monolith.

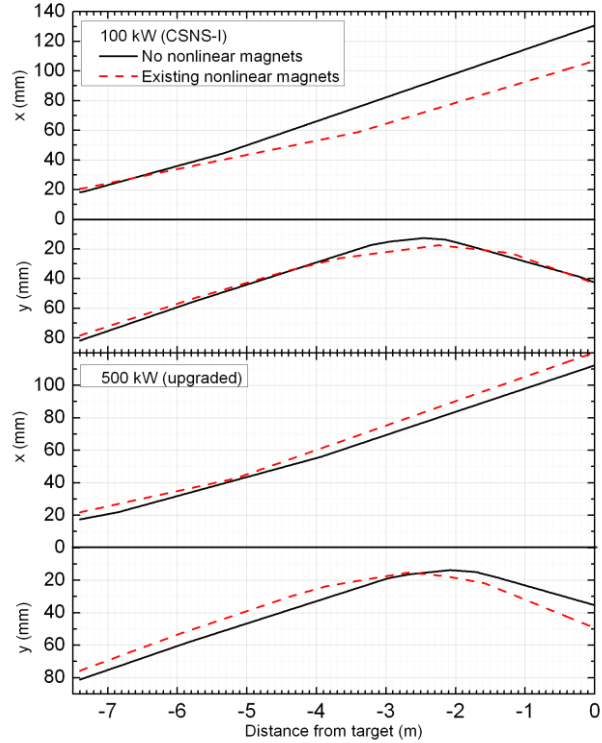

Figure 3: Incident proton beam envelopes close to the target at beam powers of 100 kW and 500 kW.

**3.2 Apertures of collimators**

For this collimation system, it must satisfy the requirements for both the proton beam and the neutron flux. For the proton beam, it will remove the very sparse halo particles in the normal operation mode and the low-power operation mode with nonlinear magnets off-line by taking into account the power limitation of 500 W in each of the collimators, and it will also absorb a part of the deviated beam in the case of RCS extraction kicker misfires. For the neutron flux, it should close the apertures as much as possible to reduce the neutron flux in the RTBT. Main parameters for the collimation system design are listed in Table 1. In the studies of the RTBT beam optics, a dual-Gaussian distribution each truncated to ±3σ with the beam core emittances of 80 πmm mrad and the beam halo emittance of 250 πmm mrad is assumed [11]. Figure 4 and Figure 5 show the beam spots on target produced by TURTLE calculations for two different operation modes. Obviously, in normal operation mode and low-power operation mode, no less than 98.0% protons have been inside the nominal footprint at the beam power of either 100 kW or 500 kW. In these cases, it is not necessary to modify the proton beam spots on the target profile by using collimators to scrape the beam halo. When it happens with the RCS extraction kicker misfires, there will be quite important beam loss in the collimators; however, this occasional beam loss does not pose problems to the collimators but as a safeguard to protect the proton beam window and target moderators.

Table 1: Proton beam parameters on the target at CSNS-I and CSNS-II.

| Proton beam parameters | CSNS-I | CSNS-II |
|---|---|---|
| Beam power | 100 kW | 500 kW |

| Energy | 1.6 GeV | 1.6 GeV |
|---|---|---|
| Current | 62.5 μA | 312.5 μA |
| Dimensions of the nominal footprint on target | 40 mm (H)×120 mm (V) | 60 mm (H)×160 mm (V) |
| Ratio of peak current density to average current density | 1.73 | 1.76 |
| Particles within footprint | > 98.3% | > 99.8% |

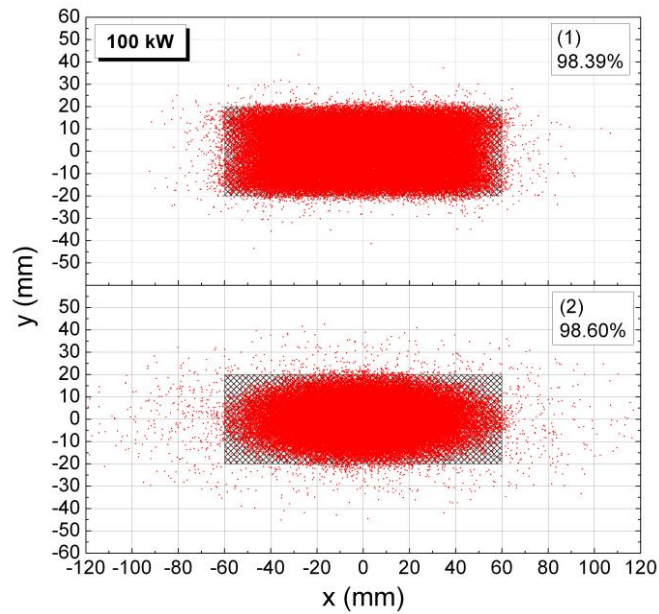

Figure 4: Beam spots on target profile in two operation modes at a power of 100 kW. The shaded rectangle is the size of nominal footprint on the target and the percentages of protons inside the footprint are given on the upright corner.

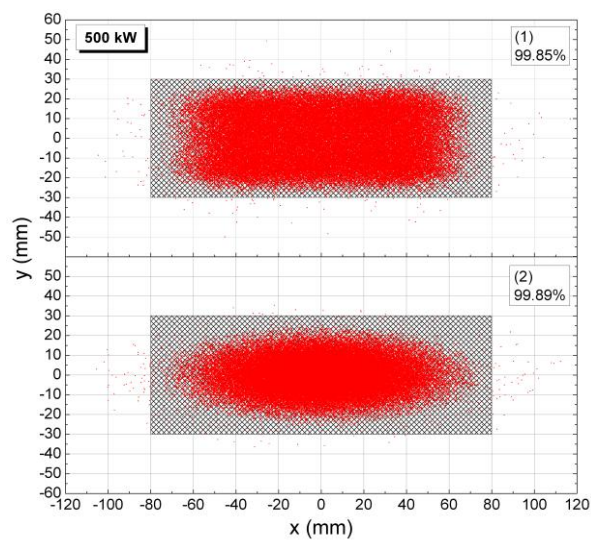

Figure 5: Same as Figure 4 but at a higher total beam power of 500 kW.

The beam dynamics studies on the RCS suggest different emittances for the beam extracted from the RCS at CSNS-I and CSNS-II, especially the core emittance. In the normal operation mode, no proton particles are allowed to lose in the collimators for both CSNS-I and CSNS-II. This can define the apertures of the collimators, which are assumed to be slanted by following the beam envelopes. Due to the similarity of apertures at CSNS-I and CSNS-II, the apertures of the collimators can be designed to be fixed ones to ease the maintenance in this high radioactive region. The apertures are listed in Table 2.

Table 2: Calculated and chosen apertures for the collimators. The chosen apertures are taken to accommodate the maximum beam envelopes at both CSNS-I and CSNS-II.

| Collimators | Collimator 1 (Horizontal waist) | Collimator 2 (Vertical waist) |
|---|---|---|
| Calculated apertures at CSNS-I | 48 mm (x) ×164 mm (y) <br> $\Delta\theta x$: 12.5mrad, $\Delta\theta y$: -2.8 mrad | 116 mm (x) ×74 mm (y) <br> $\Delta\theta x$: 16.2 mrad, $\Delta\theta y$: -1.4 mrad |
| Calculated apertures at CSNS-II | 48 mm (x) ×164 mm (y) <br> $\Delta\theta x$: 10.5 mrad, $\theta y$: -3.4 mrad | 106 mm (x) ×80 mm (y) <br> $\Delta\theta x$: 16.0 mrad, $\Delta\theta y$: -2.7 mrad |
| Chosen apertures | 48 mm (x) ×164 mm (y) <br> $\Delta\theta x$: 12.5 mrad, $\Delta\theta y$: -2.8 mrad | 116 mm (x) ×80 mm (y) <br> $\Delta\theta x$: 16.2 mrad, $\Delta\theta y$: -1.4 mrad |

### 3.3 Radiation dose rate in the RTBT tunnel

As mentioned above, the most important role of the collimators is to intercept the neutrons and other particles back-streamed from the target to reduce the radiation dose rate in the RTBT. The secondary particles produced in the target with the real target-moderator-reflector design have been studied by the CSNS target group and the authors for other purposes [3]. In this subsection, the shielding effects of the collimators are simulated by using FLUKA, assuming no proton beam losses. The rectangular beam passage in the target shielding monolith has dimensions of 250 mm ×250 mm from the PBW to the Helium container, if no specific collimators are added here. This aperture is very close to the diameter of the proton beam pipe. For comparison, the radiation dose level by the assumed maximum proton beam loss level of 1 W/m along the RTBT is also simulated. In this case, the target and its containers are set to blackbodies in FLUKA which are just used to terminate all particles trajectories including the incident proton, and avoid producing the back-streaming particles on target. In Figure 6, the ambient dose equivalent rates in this tunnel are obtained without or with collimators for the cases with collimators and without collimators, where the ICRP74 and Pelliccioni data [14] are used to perform the dose conversion. From the simulation results, we can find that the collimators are very effective in reducing the radiation dose rate in the RTBT tunnel. In addition, the only neutron contribution represents more than 98% of the total dose equivalent including other contributions from such as gamma, electrons and so on at the exit of Collimator 1. From Figure 6 (a), one can see, the dose equivalent rate induced by the back-streaming particles is one to two orders of magnitude higher than that induced by the proton beam loss.

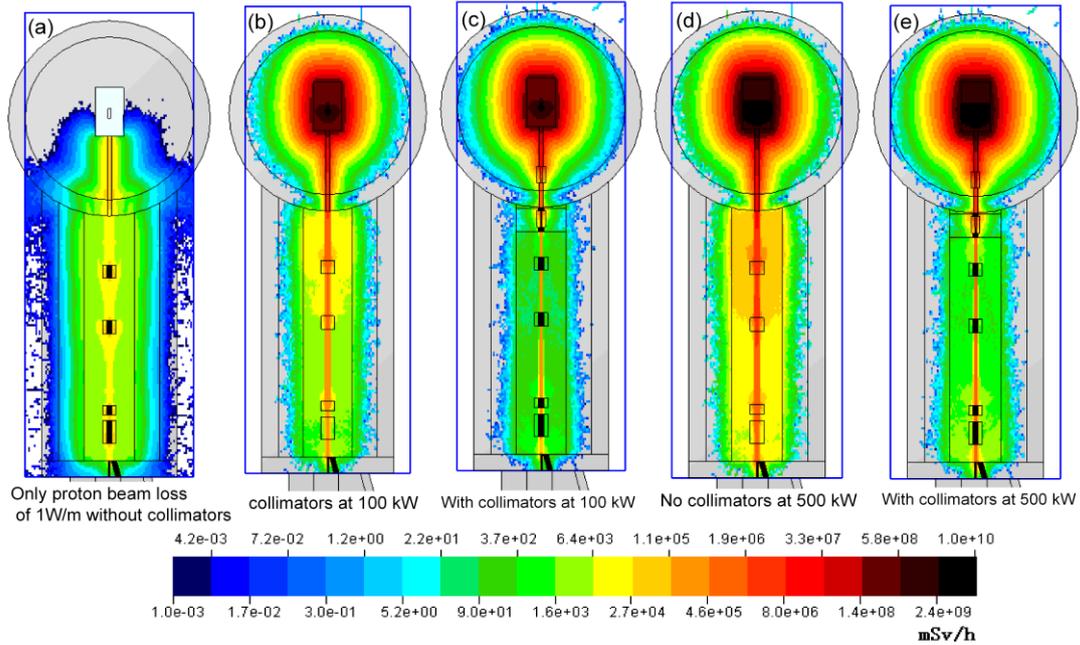

Figure 6: Dose equivalents without/with collimators in the interface region at CSNS-I and CSNS-II. The (a) depicts the dose equivalent rate in the RTBT tunnel with the only beam loss of 1W/m and the others depict that induced by the only back-streaming particles.

### 3.4 Radiation-relevant lifetimes of the last magnets in the RTBT

For the devices in the RTBT exposed to the high radiation field mainly due to the back-streaming neutrons, the conventional magnets made in epoxy-resin copper coils are especially concerned. Epoxy resin which plays a role of insulation in magnet coils will become fragile and lose insulation after suffering high-dose irradiation. Thus, it is a key factor to affect the lifetimes of the magnets in the last RTBT section. Certainly, the other devices containing organic materials in the section have similar situation.

The basic chemical elements and the portions for epoxy resin are: Carbon (76.3%), Oxygen (16.9%) and Hydrogen (6.8%). Usually, the maximum absorbed dose for epoxy resin is about $2.1 \times 10^6$ Gy. In the calculations, the annual operation time of about 5000 hours at CSNS is assumed, and the radiation dose rate in the RTBT tunnel is obtained by using the pretended polymer-alanine dosimeters (PAD) [15] in FLUKA. PAD is usually used to measure the dose rate and assess the absorbed dose of the organic materials in CERN accelerator tunnels. From Table 3, one can find that with the collimators all the five magnets in the last RTBT section can satisfy the lifetime requirements of at least thirty years at the highest beam power 500 kW. Therefore, one can ensure that MIC type magnets are not necessary at CSNS.

Table 3: Calculated radiation dose rates and lifetimes of the magnets

| No collimators | | | | |
|---|---|---|---|---|
| Distance from target | Dose rate (Gy/h) | | Lifetime (years) | |
| | 100 kW | 500 kW | 100 kW | 500 kW |
| 8.89 m (Q30D) | 17.92 | 89.59 | 23.4 | 4.7 |
| 12.31 m (Q29D) | 4.08 | 20.40 | 102.9 | 20.6 |
| 17.58 m (Q28A) | 4.05 | 20.23 | 103.8 | 20.8 |
| 18.44 m (RT_BH) | 2.22 | 11.10 | 189.2 | 37.8 |
| With collimators | | | | |
| | Dose rate (Gy/h) | | Lifetime (years) | |
| Distance from target | 100 kW | 500 kW | 100 kW | 500 kW |
| 8.89 m (Q30D) | 0.81 | 4.05 | 518.0 | 103.6 |
| 12.31 m (Q29D) | 1.22 | 6.13 | 342.4 | 68.5 |
| 17.58 m (Q28A) | 1.40 | 7.01 | 299.5 | 59.9 |
| 18.44 m (RT_BH) | 0.35 | 1.74 | 1207.7 | 241.5 |

## 4. Conclusions

A collimation system including two collimators in the CSNS accelerator-target interface region has been designed to reduce the radiation dose rate mainly induced by the back-streaming neutrons from the target. The simulations show that the collimators can reduce the ambient radiation dose rate by about two orders of magnitude and the exposure dose rates to the magnets in the last RTBT section by a factor of about 20. This can ensure that usual magnets with epoxy-resin coils can be used in the region. Certainly, the reduction in radiation dose rate by the collimators is also helpful in prolonging the lifetimes of the other less radiation-resistant devices in the region, the maintenance and the shielding design of the tunnel which is in the experimental hall. Even more, similar collimation method can be applied to other high power beam applications, such as the accelerator-target interface in Accelerator-Driven Subcritical System (ADS).

## 5. Acknowledgments

The authors want to thank the other CSNS and C-ADS colleagues for discussions. Projects 11235012 and 10975150 supported by National Natural Science Foundation of China.